\documentstyle[twocolumn,prb,aps]{revtex}
\begin{document}
\draft
\title
{Entangled Bell and GHZ states of excitons in coupled quantum dots}
\author{Luis Quiroga$^a$ and Neil F. Johnson$^b$}
\vspace{.05in}
%
\address
{$^a$Departamento de F\'{\i}sica, Universidad de Los Andes, A.A. 4976, 
Santaf\'e de Bogot\'a,  
Colombia}
\address
{$^b$Physics Department, Clarendon Laboratory, Oxford University,
Oxford OX1 3PU, U.K.}
%
\maketitle

\begin{abstract}
We show that excitons in coupled quantum dots are ideal candidates 
for reliable preparation of entangled states in solid-state systems. An
optically controlled exciton transfer process is shown to lead to the
generation of
Bell and GHZ states in systems comprising two and three coupled dots,
respectively.
The strength and duration of selective light-pulses for
producing maximally entangled states are identified by both analytic, and
full numerical,
solution of the quantum dynamical equations. Experimental
requirements to build such entangled  states are discussed.
\end{abstract}
\vspace{.25in}


\narrowtext

Quantum information, quantum computation, quantum
cryptography and quantum teleportation represent exciting new arenas which
exploit intrinsic quantum mechanical correlations.\cite{PW} A fundamental
requirement for the experimental realization of such proposals is the
successful
generation of highly entangled quantum states. In particular, coherent
evolution
of two quantum bits (qbits) in an entangled state of the Bell type is
fundamental
to both quantum cryptography and quantum teleportation. Maximally entangled
states of three qbits, such as the so-called Greenberger-Horne-Zeilinger
(GHZ) states\cite{AJP}, are not only of intrinsic interest but are also of
great practical importance in such proposals. New systems and
methods for the preparation and measurement of such maximally  entangled
states
are therefore being sought intensively. Most of the theoretical and
experimental activity to date has been associated with atomic and
quantum-optic systems
\cite{Cirac1,Cirac2,molmer}. 

Solid-state realizations of such quantum-based phenomena have received
little attention despite the fact that semiconductor nanostructures
such as quantum dots (QDs), with quantum-mechanical electron  confinement in
all
three directions, have been fabricated and studied by many
groups.\cite{Neil} 
In addition, recent experimental work by Bonadeo et al.
\cite{Gammon1,Gammon2}
suggests
that optically-generated electron-hole pairs (excitons) in semiconductor QDs
represent ideal candidates for achieving coherent wavefunction control on
the
nanometer and femtosecond scales.

In this paper we give a detailed prescription for producing
such entangled states in semiconductor quantum dot systems. We
show that the resonant transfer interaction between spatially separated 
excitons can be exploited to produce such entanglement, starting from
suitably initialized states. The system requirements are realizable in
current experiments employing ultrafast optical
spectroscopy of quantum dots. 

When two quantum dots are sufficiently close, 
there is a resonant energy-transfer process originating from the Coulomb
interaction whereby an exciton can hop between dots\cite{Mahler}.
Experimental
evidence of such energy-transfers between quantum dots was reported
recently\cite{Gammon1}; the resonant process also plays a fundamental role
in
biological and organic systems, and is commonly called the Forster
process\cite{PT}. Unlike usual single-particle transport measurements, the
Forster process does not require the physical transfer of the electron and
hole,
just their energy. Hence it is relatively insensitive to effects of
impurities
which lie between the dots. Consider a system of $N$ ($N=2,3$ or $4$)
identical, equispaced
QDs containing no net charge, radiated by long-wavelength classical light.
Ignoring
any constant energy terms, the following Hamiltonian describes the formation
of
single excitons within the individual QDs and their inter-dot hopping: 
\begin{eqnarray}
\nonumber
H(t)=\frac {\epsilon}{2} \sum_{n=1}^{N}{(
e_{n}^{\dag}e_{n}-h_nh_n^{\dag})}
- \frac {1}{2} V \sum_{n,n'}^{N}\big(
e_{n}^{\dag}h_{n'}
e_{n'}h_{n}^{\dag}\\
+ h_{n}e_{n'}^{\dag}
h_{n'}^{\dag}e_{n}
\big) + 
E(t) \sum_{n=1}^{N}{e_{n}^{\dag}h_{n}^{\dag}} +
E^*(t) \sum_{n=1}^{N}{h_{n}e_{n}}\ \ .
\end{eqnarray}
Here $e_{n}^{\dag}$ ($h_n^{\dag}$) is the electron (hole) creation operator
in 
the
$n$'th QD. The 
QD band gap is $\epsilon$ while $V$ represents the
interdot Coulomb interaction and hence the Forster process. 
The dots are equidistant
from each other, i.e. $N=2$ dots on a line, $N=3$ dots at the vertices of an
equilateral triangle, $N=4$ dots at the
vertices of a pyramid; hence $V$ is not a function of $n$ or $n'$. 
The time-dependence of $E(t)$ describes the pulse shape, while the magnitude
includes
the electron-photon coupling and the incident electric field strength. The
Hamiltonian can be easily manipulated using quasi-spin operators
\begin{eqnarray}
\nonumber
J_+=\sum_{n=1}^{N} {e_{n}^{\dag}h_{n}^{\dag}}\ ; 
\ \ J_-=\sum_{n=1}^{N}{h_{n}e_{n}}\ ;\\
J_z=\frac {1}{2} \sum_{n=1}^{N} {(e_{n}^{\dag}e_{n}-
h_{n}h_{n}^{\dag})}\ \ .
\end{eqnarray}
These satisfy the usual angular momentum commutation relationships:
$[ J_+,J_-]=2J_z$, 
$[ J_{\pm},J_z]=\mp J_{\pm}$.
The Hamiltonian now takes the form
\begin{eqnarray}
H(t)=\epsilon J_z-V(J^2-J_z^2)+E(t)J_++E^*(t)J_-\ \ .
\end{eqnarray}
$H(t)$ contains a non-linear term that can be exploited to generate
entangled states.

We consider a 
rectangular radiation pulse, starting at time
$t=0$ with central frequency $\omega$, given by
$E(t)={\it A} {\rm cos}(\omega t)$: such a pulse is relatively
straightforward to
achieve experimentally. The time evolution of any initial state under the
action
of $H$ in Eq.(3) is
easily performed by means of the pseudo 1/2-spin operator 
formalism.\cite{wokaun,vega} Single transition operators are
defined by
\begin{eqnarray}
\nonumber
\langle i|J_x^{r-s}|j\rangle =\frac {1}{2} (\delta_{ir}\delta_{js}+
\delta_{is}\delta_{jr})\\
\nonumber
\langle i|J_y^{r-s}|j\rangle =\frac {i}{2} (-\delta_{ir}\delta_{js}+
\delta_{is}\delta_{jr})\\
\langle i|J_z^{r-s}|j\rangle =\frac {1}{2} (\delta_{ir}\delta_{jr}-
\delta_{is}\delta_{js})
\end{eqnarray}
where $r-s$ denotes the transition between states $|r\rangle$ and
$|s\rangle$
within a  given $J$ subspace. The
three operators belonging to one particular transition $r-s$ obey standard
angular momentum commutation relationships
$\left [ J_{\alpha}^{r-s},J_{\beta}^{r-s} \right ]=iJ_{\gamma}^{r-s}$
where $(\alpha,\beta,\gamma)$ represents a cyclic permutation of
$(x,y,z)$. Note that operators belonging to non-connected transitions 
commute:
$\left [ J_{\alpha}^{r-s},J_{\beta}^{t-u} \right ]=0$
with $\alpha, \beta=x,y$ or $z$.
In order to gain physical insight into the dynamics of such a multi-exciton
problem, some approximations are necessary: a common assumption, valid 
when $\epsilon\gg V$, is the so-called
rotating wave approximation 
$U=e^{-i\omega J_zt}$.
Suppressing rapidly
oscillating terms, the Hamiltonian in the rotating frame (RF) becomes 
\begin{eqnarray}
H_r=U^{\dag}HU+i\frac {dU^{\dag}}{dt}U=\Delta \omega J_z-AJ_x-V(J^2-J_z^2)
\end{eqnarray}
where $\Delta \omega=\epsilon - \omega$ denotes the off-resonance condition.
We now
show that this Hamiltonian leads to the generation of entangled states
from suitably initialized states.

{\it Two coupled QDs: Bell states.} Here we describe the
light excitation procedure to obtain a maximally entangled Bell state of the
form $|\Psi\rangle=|00\rangle+e^{i\phi}|11\rangle$
with 0 (1) denoting a zero-exciton (single exciton) QD. The phase angle
$\phi$ can be arbitrary.
In order to highlight the physical aspects of the procedure, we first derive
an
approximate analytical solution of the dynamical
equation
governing the system's matrix density. Starting with the initial condition 
representing
the vacuum of excitons, $|J=1,M=-1\rangle$, only the $J=1$ subspace is
optically active while the $J=0$ subspace remains dark. 
We choose the basis of eigenstates of $J^2$ and $J_z$,
$\{
|0\rangle=|J=1,M=-1\rangle$, $|1\rangle=|J=1,M=0\rangle$,
$|2\rangle=|J=1,M=1\rangle\}$, as an appropriate representation for this
problem. $|0\rangle$ represents the vacuum for excitons,
$|1\rangle$ denotes a symmetric delocalized single-exciton state while
$|2\rangle$ represents the biexciton state. The RF-Hamiltonian and initial
density  matrix can be expressed in terms of pseudo-spin operators as
follows:
\begin{eqnarray}
\nonumber
H_r=-2\Delta \omega J_z^{0-2}+\frac {2V}{3}(J_z^{0-1}-J_z^{1-2})\\
-\sqrt{2} A (J_x^{0-1}+J_x^{1-2})\ \ ,
\end{eqnarray}
\begin{eqnarray}
\rho(0)=\frac {1}{3}I+\frac {2}{3}(J_z^{0-1}+J_z^{0-2})\ \ .
\end{eqnarray}
Here $I$ denotes the identity matrix in the subspace $J=1$.

In the absence of light, the energy levels of the system are given
by $E=-\Delta \omega-V$, $-2V$ and $\Delta \omega-V$. Note that the energy
separation
between states $|0\rangle$ and $|2\rangle$ is unaffected by the inter-dot
interaction  $V$. Now we consider the action of a  pulse of light at
resonance, i.e. $\Delta \omega=0$, and amplitude given by $A\ll V$. We
assume that the
decoherence processes are negligibly small on the time scale of the
evolution (see later).
The density matrix at time $t$ becomes
\begin{eqnarray}
\nonumber
\rho (t)=\frac {1}{3}I+\left [ {\rm cos}(\omega_2t)+\frac {1}{3}\right ]
J_z^{0-1}\\
+
\left [ {\rm cos}(\omega_2t)-\frac {1}{3}\right ] J_z^{1-2}
-{\rm sin}(\omega_2t)J_y^{0-2}
\end{eqnarray}
which exhibits the generation of coherence between vacuum and
biexciton
states through the operator $J_y^{0-2}$, which oscillates at
frequency $\omega_2=A^2/V$.
\par The Bell state $|\Psi\rangle=|00\rangle+e^{i\phi}|11\rangle$ has a
corresponding density matrix
$\rho_{Bell}=I/3+J_z^{0-1}/3-J_z^{1-2}/3+{\rm cos}(\phi)J_x^{0-2}-
{\rm sin}(\phi)J_y^{0-2}$.
Comparing this last equation with Eq.(8), we see that the system's quantum
state
at time $\tau_2=\pi V/2A^2$ corresponds to a maximally entangled Bell state
of
the desired
form with $\phi=\pi/2$. The time evolutions of populations and 
coherences for an initial vacuum state are plotted in Fig. 1. The evolution
of populations of the vacuum $\rho_{00}$ and the biexciton $\rho_{22}$ 
states are shown in Fig. 1a.
Clearly our approximate analytic calculation describes the
system's evolution very well when compared with the exact
numerical  solution (Fig. 1a). Figure 1b shows the 
overlap, $O(t)=Tr \left [ \rho_{Bell}\rho(t) \right ]$, 
between the maximally entangled Bell state and the one
obtained by applying a rectangular pulse of light at resonance.
The thick solid line (Fig. 1b) describes $O(t)$ with a maximally entangled
Bell
state in the rotating frame, while the thin solid line (Fig. 1b) represents
the
overlap with a Bell state transformed to the laboratory frame: obviously the
RF
case corresponds to the amplitude evolution of the  laboratory frame signal.
The dashed line illustrates the approximate solution overlap in
the RF.  The approximate solution works very well,
supporting the idea that a selective Bell pulse of length
$\tau_2=\pi V/2A^2$ can be used to create a Bell state ($\phi=\pi/2$) in
the system of two coupled QDs. 
The same conclusion can also be drawn from the time evolution
of the overlap between the exact Bell-state density matrix and the one
obtained
directly from the numerical calculation. Therefore, 
the existence of a selective Bell pulse is numerically confirmed.

{\it Three coupled QDs: GHZ states.}
Next consider three
quantum dots of equal size placed at the corners of an equilateral triangle.
We can consider the $J=3/2$ subspace 
as being the only optically active subspace: the other two $J=1/2$ subspaces
remain
optically dark. We work in the basis set $|J=3/2,M\rangle$,
$\{|0\rangle=|3/2,-3/2\rangle$, $|1\rangle=|3/2,-1/2\rangle$,
$|2\rangle=|3/2,1/2\rangle$, $|3\rangle=|3/2,3/2\rangle\}$, where
$|0\rangle$ is
the vacuum state, $|1\rangle$ is the symmetric delocalized single-exciton
state,
$|2\rangle$ is the symmetric delocalized biexciton state and $|3\rangle$ is
the
triexciton state. In the absence of radiation the energy levels
are
$-3(\Delta \omega+V)/2, -(\Delta \omega+7V)/2, (\Delta \omega-7V)/2$ and
$3(\Delta \omega-V)/2$. 
In terms of pseudo-spin operators, the Hamiltonian in RF, including the 
radiation term, is now given by
\begin{eqnarray}
H_r=-\Delta \omega (3J_z^{0-3}+J_z^{1-2})+2V(J_z^{0-1}-2J_z^{2-3})\\
\nonumber
-A \left [ \sqrt{3} (J_x^{0-1}+J_x^{2-3})+2J_x^{1-2} \right ]
\end{eqnarray}

Two kinds of maximally entangled GHZ states have to be 
considered\cite{pan}:\\

\noindent (i) the entangled state between vacuum and triexciton states given
by $|GHZ\rangle_1=\frac {1}{\sqrt{2}}(|0\rangle+e^{i\phi}|3\rangle)$ or in
terms
of its associated density matrix by 
$\rho_{G1}=I/4+J_z^{0-1}/2
-J_z^{2-3}/2+{\rm cos}(\phi)J_x^{0-3}+{\rm sin}(\phi)J_y^{0-3}$, where $I$
denotes the identity matrix in the $J=3/2$ subspace. We now show that
this state can be generated after an appropriate $\pi/2$-pulse. Starting
with a zero-exciton state $|0\rangle$, the evolved state under the action of
Hamiltonian Eq.(9) at resonance, i.e. $\Delta \omega=0$, 
can
be obtained in a straighforward way in the limit $A/V\ll 1$ using the
properties of
pseudo-spin operators:
\begin{eqnarray}
\nonumber
\rho(t)=\frac {1}{4} I+\left [ {\rm cos}(\omega_3t)+\frac {1}{2}\right ]
J_z^{0-1}+
{\rm cos}(\omega_3t)J_z^{1-2}\\
+\left [ {\rm cos}(\omega_3t)-\frac {1}{2}\right ] J_z^{2-3}
+{\rm sin}(\omega_3 t) J_y^{0-3}
\end{eqnarray}
with $\omega_3=d_--d_++A$ and 
$d_\pm=V \left [ 1\pm\frac {A}{V}+(\frac {A}{V})^2 \right] ^{1/2}$.
Clearly $|GHZ\rangle_1$ (with $\phi=\pi/2$) can be generated with 
a $\pi/2$-pulse of length $\tau_3=4\pi V^2/3A^3$.
In Fig.2a we show the overlap between the exact density matrix
and that corresponding to state $|GHZ\rangle_1$. The dashed line
shows the overlap using our approximate density matrix, Eq.(10).
\\ 

\noindent (ii) the entangled state between a single exciton $|1\rangle$
and the  biexciton $|2\rangle$ given by $|GHZ\rangle_2=\frac
{1}{\sqrt{2}}(|1\rangle+e^{i\phi}|2\rangle)$, or in terms of the
corresponding density
matrix
$\rho_{G2}=I/4-J_z^{0-1}/2+J_z^{2-3}/2+{\rm cos}(\phi)J_z^{1-2}+
{\rm sin}(\phi)J_y^{1-2}$. In order to generate $|GHZ\rangle_2$ the initial
condition 
must be modified.
Using a suitably designed preparation sequence of
pulses a new initial state, corresponding to a single exciton state
$|1\rangle$, 
can be generated. Evolution of this new initial state
under $H_r$ (Eq.(9)) with $\Delta \omega=0$,
generates a density matrix at time $t$ given by
\begin{eqnarray}
\nonumber
\rho(t)=\frac {1}{4} I-\frac {1}{2}J_z^{0-1}
+{\rm cos}(\omega_3^{\prime}t)J_z^{1-2}+\frac {1}{2}J_z^{2-3}+\\
{\rm sin}(\omega_3^{\prime}t)J_y^{1-2}
\end{eqnarray}
where $\omega_3^{\prime}=d_--d_+-A$. Comparing this last result with the
density matrix corresponding to a $|GHZ\rangle_2$ state we see that a
$\pi/2$-pulse can be identified with a duration given by
$\tau_3^{\prime}=\pi/4A$. Figure 2b shows the overlap between $\rho(t)$
(Eq.(11)) 
and  $\rho_{G2}$. There is good agreement between
our approximate result, Eq.(11), and the exact one. 

We emphasize that the two maximally entangled GHZ states
considered above have very
different frequencies.
In the limit
$A/V\ll 1$ the state $|GHZ\rangle_1$ oscillates at the frequency 
$\omega_3\sim 3A^3/8V^2$ while
the state $|GHZ\rangle_2$ oscillates at the larger frequency 
$\omega_3^{\prime}\sim 2A$. This feature
should enable each of
these maximally entangled $GHZ$ states to be manipulated separately in
actual
experiments, even if the initial state is mixed. Furthermore we note
that, after the preparation step, the system is evolving under the action
of the Hamiltonian in Eq. (5) with
$\Delta\omega=A=0$: each one of the maximally entangled states
discussed in this work are eigenstates of this remaining
Hamiltonian.  Hence in the laboratory frame
 $|Bell\rangle$ oscillates at frequency $2\epsilon$,
$|GHZ\rangle_1$ oscillates at frequency $3\epsilon$ and
$|GHZ\rangle_2$ oscillates at frequency $\epsilon$.

Experimental observation of these Bell and GHZ states should be possible
with present ultrafast semiconductor optical
techniques\cite{Gammon1,Gammon2}. For
instance,  for GaAs QDs $\epsilon=1.4 {\rm eV}$ which implies a resonance
optical
frequency $\omega=2$x$10^{15} {\rm sec}^{-1}$. From the results above,
it follows that in order to generate maximally entangled exciton states,
$\pi/2$-pulses with sub-picosecond duration 
should be used. Femtosecond optical spectroscopy in GaAs-based 
nanostructures is now currently in use\cite{Gammon1,Gammon2}. On the
other hand, wide-gap semiconductor QDs, like ZnSe-based QDs, should do
better because of the shorter required $\pi/2$-pulse length: $\epsilon=2.8
{\rm eV}$
which leads to an optical resonance frequency of $\omega=4$x$10^{15}
{\rm sec}^{-1}$.
Femtosecond spectroscopy is also currently available for this 
system.\cite{axt} In addition, the corresponding increase in the effective gap
will yield a larger exciton binding energy: typical decoherence 
mechanisms (e.g. acoustic phonon
scattering) will hence become less effective.
A surprising conclusion of our results is that entangled-state preparation is
facilitated by {\em weak} light fields (i.e. $A\ll V$): strong fields cause
excessive
oscillatory behavior in the density matrix. This paper 
has considered the relatively straightforward experimental situation of {\it
global}
excitation pulses, i.e. pulses acting simultaneously on the entire QD
system.
However, other possibilities exist such as  near-field optical 
spectroscopy\cite{chavez} which allows the optical excitation and detection
of 
{\it individual} QD signals. In this way, maximally entangled states with
different
symmetries  can be obtained.

\par Finally there is the well-known but difficult problem of 
decoherence. Existing solid-state proposals for quantum computers include
quantum gates
in coupled QDs based on electron spin effects\cite{Burkard} and electronic
charge
effects\cite{Barenco}. Although electronic charge effects are subject to
phonon
decoherence,  subdecoherent information may still be encoded in such a
quantum-dot
array as described recently by Zanardi et al.\cite{Zanardi} 
The present paper has focussed on exciton-based systems, since these have
already been shown to exhibit good coherence properties up to the picosecond
time-scale\cite{Gammon2}. Phonon decoherence will therefore be relatively
unimportant on
this time-scale. In addition, since no inter-dot transport of particles
occurs, 
scattering due to impurities etc. lying between the dots will be negligible.
A
detailed analysis of all possible decoherence times is beyond the scope of
the present
work, but will be addressed elsewhere. 

\par In summary, we have shown how maximally entangled Bell and GHZ states 
can be generated using the optically driven resonant transfer
of excitons between quantum dots. Selective Bell and GHZ pulses have been 
identified
by an approximate, yet accurate, analytical approach which should prove a
useful
tool when designing experiments. Exact numerical calculations 
confirm the existence of such
$\pi/2$-pulses for the generation of maximally entangled states in coupled
dot systems. 

L.Q. acknowledges to Ferney J. Rodr\'{\i}guez for helpful discussions.
The authors acknowledge support from COLCIENCIAS project No.1204-05-264-94.
N.F.J. also
thanks the EPSRC (UK).

\newpage
\centerline{\bf Figure Captions}

\bigskip

\noindent Figure 1: (a) Population of the vacuum state $\rho_{00}$ and
biexciton state $\rho_{22}$ in two coupled QDs, as a function of time.
(b) Time-evolution of overlap with maximally entangled Bell state.
$\epsilon=1$, 
$V=\epsilon/10$ and $A=V/5$. Thin solid line shows exact numerical result in
the laboratory frame. Thick solid line in (b) represents the exact
numerical solution in the rotating frame. Dashed line shows approximate
analytic
result.

\bigskip

\noindent Figure 2: Time-evolution of overlap with maximally
entangled $GHZ$ states. (a) $|GHZ\rangle_1$ and (b) $|GHZ\rangle_2$  under
the
action of a rectangular pulse of light at resonance. $\epsilon=1$, 
$V=\epsilon/10$ and $A=2V/5$. Thick solid line represents exact numerical
solution. Dashed line shows approximate analytic result.

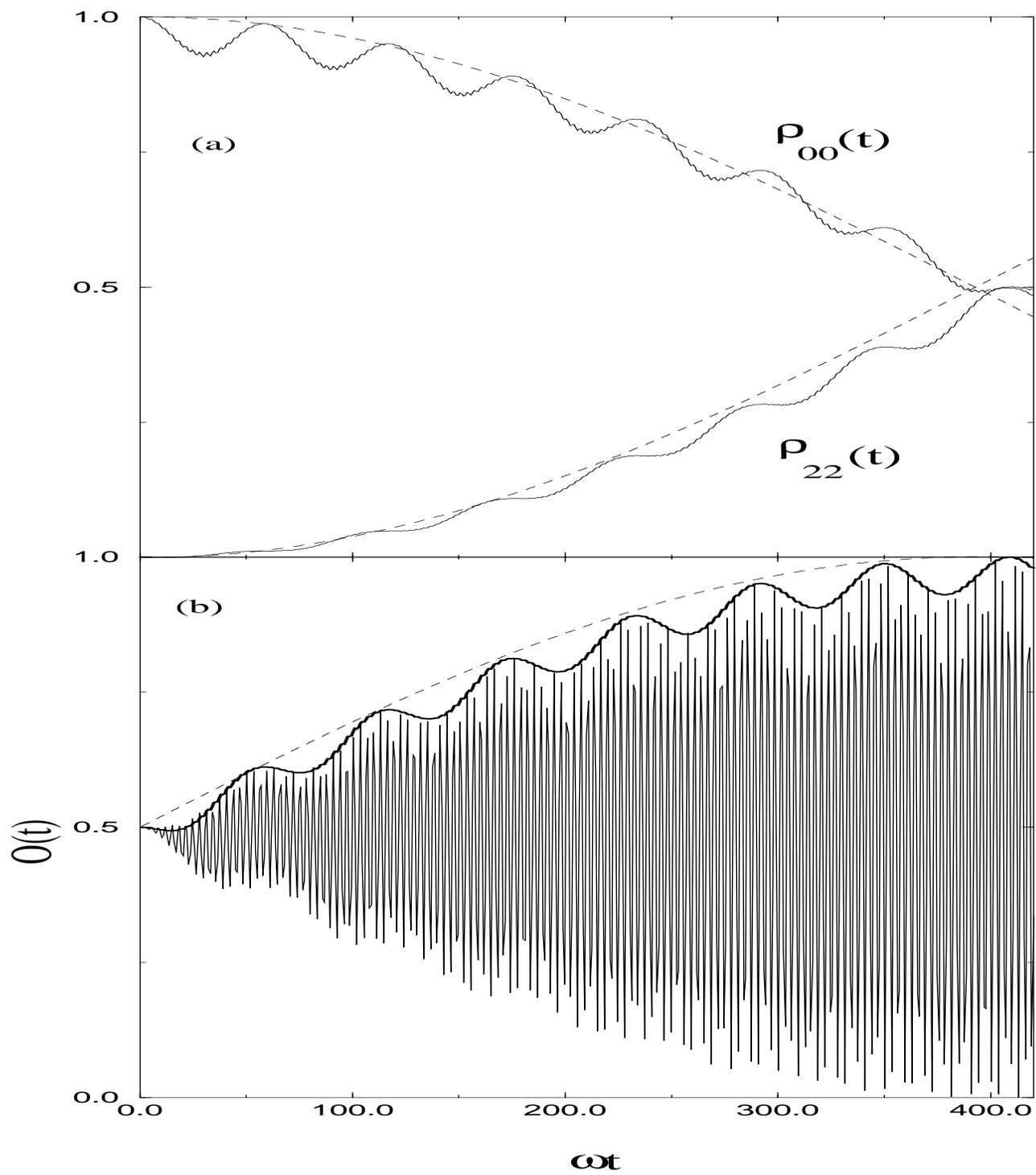
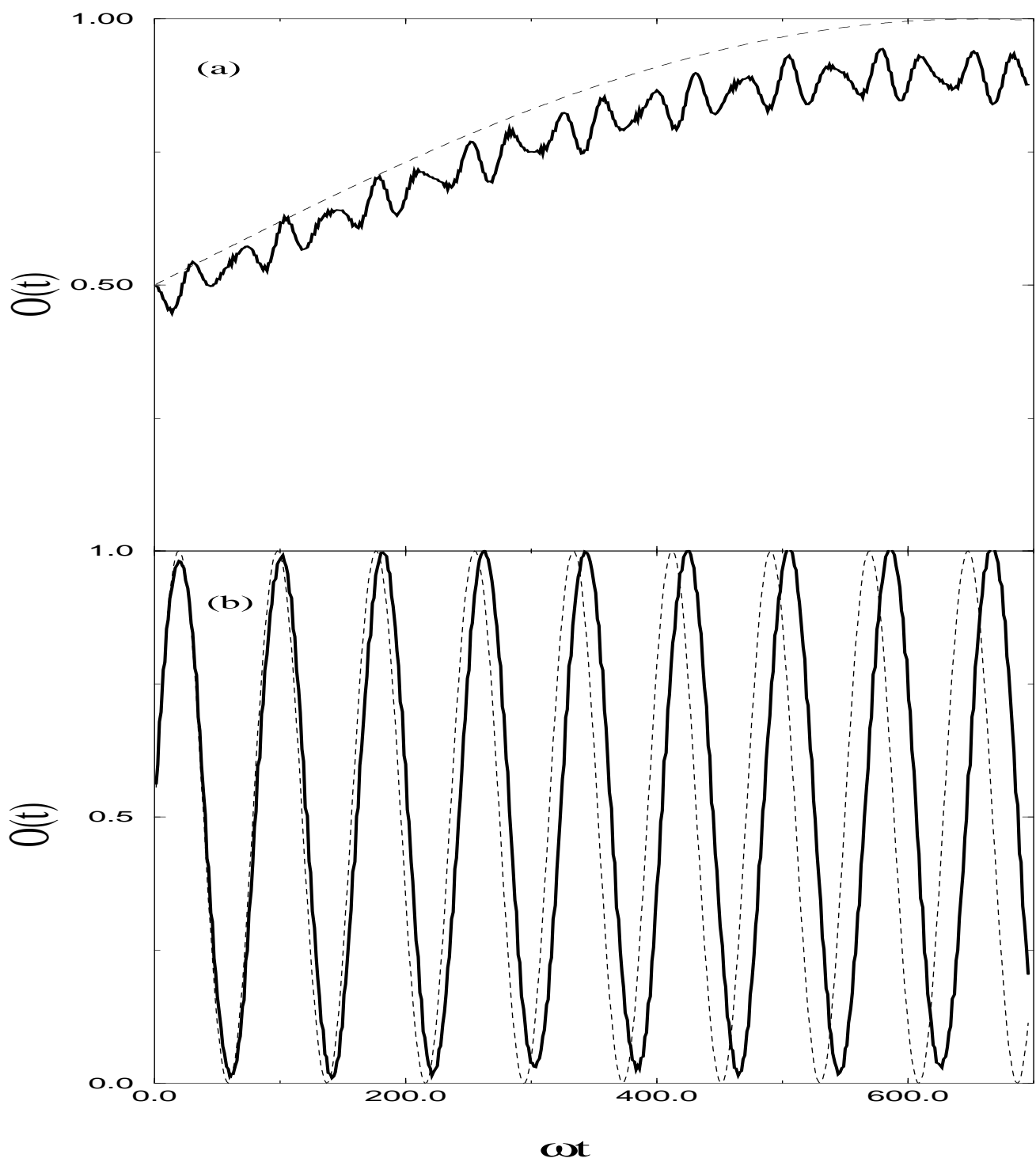
\end{document}